\documentclass[a4paper, 12pt]{article}
\usepackage{amsmath}
\usepackage{graphicx}
\usepackage{amssymb}
 \usepackage{amsthm}
\usepackage{tabularx}
\usepackage{hyperref}
\usepackage{graphicx}
\usepackage{subfigure}
\usepackage{booktabs} 
\usepackage[hang, small,labelfont=bf,up,textfont=it,up]{caption}
\usepackage{rotating}
\usepackage{relsize}
\usepackage{setspace}

\usepackage[a4paper,top=2.5cm,bottom=3cm,left=3cm,right=3cm]{geometry}
\begin{document}

\begin{center}
\textbf{Coefficient of thermal expansion of nanostructured tungsten based coatings assessed by substrate curvature method}
\newline
E. Besozzi$^a$, D. Dellasega$^{a,b}$, A. Pezzoli$^{a}$, A. Mantegazza$^{a}$, M. Passoni$^{a,b}$, M.G. Beghi$^a$
\vspace{0.5cm}
\newline
\textit{\small{
$^a$ Dipartimento di Energia, Politecnico di Milano, via Ponzio 34/3, I-20133, Milano,Italy
\newline
$^b$ Associazione EURATOM-ENEA, IFP-CNR, Via R. Cozzi 53, 20125 Milano, Italy}}
\end{center}
\begin{abstract}
The in plane coefficient of thermal expansion (CTE) and the residual stress of nanostructured W based coatings are extensively investigated. The CTE and the residual stresses are derived by means of an optimized ad-hoc developed experimental setup based on the detection of the substrate curvature by a laser system. The nanostructured coatings are deposited by Pulsed Laser Deposition. Thanks to its versatility, nanocrystalline W metallic coatings, ultra-nano-crystalline pure W and W-Tantalum coatings and amorphous-like W coatings are obtained. The correlation between the nanostructure, the residual stress and the CTE of the coatings are thus elucidated. We find that all the samples show a compressive state of stress that decreases as the structure goes from columnar nanocrystalline to amorphous-like. The CTE of all the coatings is higher than the one of the corresponding bulk W form. In particular, as the grain size shrinks, the CTE increases from 5.1 10$^{-6}$ K$^{-1}$ for nanocrystalline W to 6.6 10$^{-6}$ K$^{-1}$ in the ultra-nano-crystalline region. When dealing with amorphous W, the further increase of the CTE is attributed to a higher porosity degree of the samples. The CTE trend is also investigated as function of materials stiffness. In this case, as W coatings become softer, the easier they thermally expand.     
\end{abstract}
Tungsten coatings, coefficient of thermal expansion, substrate curvature, Pulsed Laser Deposition, residual stresses, nanostructure

\section{INTRODUCTION}

\label{Introduction}

It is well known that coatings are subjected to \textit{residual stresses}, already present at the end of the deposition process. These stresses have two main origins. \textit{Intrinsic stresses}, which are due to the deposition process itself, depending on the deposition conditions and by the mismatch of the properties between the coating and the substrate materials (e.g. lattice parameter) \cite{Marques1998}. \textit{Thermal stresses}, due to a thermal expansion mismatch between the coating and the substrate, they depend on the elastic properties of the deposited and the base material, usually rising when the sample is cooled down to room temperature after deposition. When the coated components operate at variable temperatures, additional thermal stresses generate. These stresses, typically intensifying at the film-substrate interface, can lead to coating failure, by either cracking or delamination. Predicting and monitoring these stresses is crucial to guarantee the operational integrity of the coated devices. This requires the knowledge of the elastic moduli and the CTE of the materials. In the case of coatings this cannot be taken for granted. Firstly, because the
thermomechanical properties, which depend on the specific film structure and morphology, can be significantly different from the ones of the corresponding bulk form, and depend on the deposition process. Secondly, because, 
for coatings, direct measurement can be a challenging task. 
\newline
A wide range of techniques is available to investigate the elastic properties of coatings;
namely, nanoindentation \cite{Ferrè2013, UlHamid} and various acoustic based techniques \cite{BeghiBook}, 
including Brillouin spectroscopy \cite{Besozzi2016, Sumanya2017, Ozsdolay2017, Faurie2017}, while 
little is known about the CTE of films. The standard techniques adopted to measure the CTE of
bulk materials (e.g. dilatometry \cite{Huang2005, Jackson2016}) are usually not viable for 
coatings. Several unconventional techniques have been proposed, such as X-ray diffraction 
\cite{Zoo2006, Bartosik2017, Lei2017}, ellipsometry \cite{Singh2004} and different optical based
techniques \cite{James2001}. Among them, the optical implementation of the substrate
curvature (SC) technique has shown to be one of the most promising methods 
\cite{Lei2017, Hang2010, Lima1999, Woehrl2009, Knepper2007, Dutta2016}. This method exploits 
 laser beams to detect changes in the curvature radius of the coating-substrate system upon temperature 
variations \cite{Chason2001}. 
The CTE of the coating can be then deduced if the CTE of the substrate and the elastic
properties of both the film and the substrate are known (see section 2). 
\newline
In this work, we investigate the CTE and the residual stresses of tungsten (W) coatings deposited by Pulsed Laser Deposition (PLD). W coatings are of particular interest in a wide range of technological applications, such as in microelectronic and optoelectronic devices, as absorption layers in X-ray lithography \cite{Chen2005, Radic2004, Kobayashi}, and in nuclear fusion energy \cite{Boucalossi2014, Ruset2013, Guilhem2016}. Thanks to the high versatility of PLD in tuning many process parameters (e.g. background gas pressure during deposition, laser fluence on target), both mono-elemental and multi-elemental coatings can be grown with tailored nanostructure, from amorphous to nanocrystalline, and morphology, from porous to compact \cite{Besozzi2016, Pezzoli2015,  Luo2017, Li2015}. Here, we focus on W-based coatings with three different nanostructures, namely (i) nanocrystalline W (n-W), (ii) ultra-nano-crystalline W (u-n-W) and (iii) amorphous-like W (a-W), with the aim of highlighting the correlation between the thermal expansion behavior, the residual stresses and the structural properties of the materials.  
\newline
For the coating characterization, we develop an optimized SC setup that allows the CTE determination over a wide range of temperatures (25 - 1000 $^{\circ }$C). An ad-hoc designed vacuum chamber is equipped with an optical system that drives a 2D pattern of parallel laser beams on the surface of the coated substrate, and detects the reflected beams by a CMOS sensor. The beam positions, when the sample is thermally bent, allow the direct determination of the
substrate-coating curvature as function of temperature. From curvature measurements, the residual stresses and the CTE of the coatings are derived, under the Stoney approximation \cite{Stoney}, for known elastic moduli, which have already been measured by Brillouin spectroscopy (BS) \cite{Besozzi2016}.


\section{The principle of obtaining the residual stress and the CTE of the coatings}

Upon a temperature variation, the mismatch in the CTE between the coating and the substrate, combined with the dilation constraint represented by the
film adhesion to the substrate, leads the sample to a progressive bending. The total bending depends on the difference between the CTEs of the two
materials, on their thicknesses and their elastic moduli, and obviously on the temperature itself; it is well described by the continuum mechanics
theory for multilayers \cite{2002Hsueh}. In the case of a bilayer formed by a film much thinner than the substrate, such that the stress within the film can be taken as approximately uniform, the stress within the coating can be expressed in terms of the bending curvature radius $R$ as: 
\begin{equation}
\sigma _{f}(T)=\frac{E_{s}}{1-\nu _{s}}\frac{t_{f}}{t_{s}^{2}}\frac{1}{6}(%
\frac{1}{R(T)}-\frac{1}{R_{0}})  
\label{eq1}
\end{equation}
In Eq.\ref{eq1} the sub-indexes $s$ and $f$ stand for substrate and film respectively, $t$ is the thickness, $R(T)$ the radius of curvature at temperature $T$ and $R_{0}$ the initial radius of curvature at a reference temperature. $E$ is the Young modulus and $\nu$ the Poisson's ratio. Eq. \ref{eq1} is often known as Stoney's equation \cite{Stoney, Janssen2009}. If the total film stress is only due to the thermal component, it is given by: 
\begin{equation}
\sigma _{f}=\sigma _{thermal}=\frac{E_{f}}{1-\nu _{f}}(CTE_{f}-CTE_{s})%
\Delta T\quad ,  
\label{eq2}
\end{equation}%
and, taking the derivative of eq. \ref{eq2} over temperature: 
\begin{equation}
\frac{d\sigma _{f}}{dT}=\frac{E_{f}}{1-\nu _{f}}(CTE_{f}-CTE_{s})\quad ,
\label{eqboh}
\end{equation}%
such that:%
\begin{equation}
CTE_{f}=CTE_{s}+\frac{d\sigma _{f}}{dT}\frac{1-\nu _{f}}{E_{f}} 
\label{eqCTE}
\end{equation}
\begin{figure}[!t]
\centering
\includegraphics[width = 0.6\columnwidth]{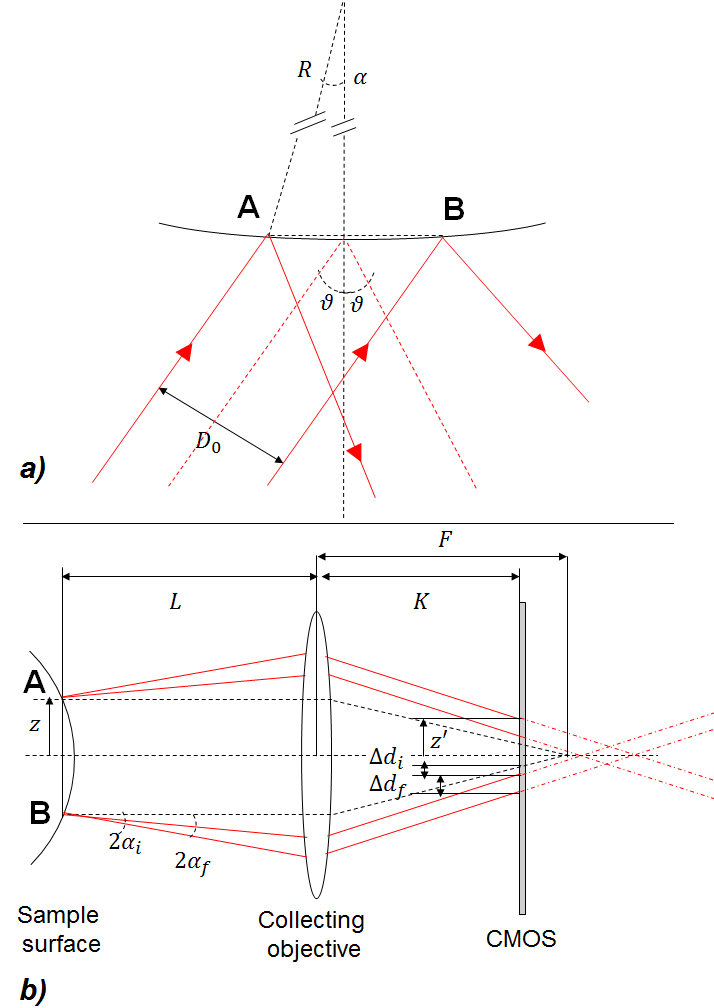}
\caption{a) Schematic principle of two initially parallel laser beams that are reflected by a curved surface. b) the reflected beams are collected by the collecting objective and recorded by the CMOS}
\label{Fig4}
\end{figure}
Once the film stress is derived from the curvature measurement by eq. \ref{eq1}, eq. \ref{eqCTE} is exploited to derive the CTE of the coatings. Equations \ref{eq1}-\ref{eqCTE}, are valid only if the elastic moduli are considered as temperature independent. A more realistic approach would clearly consider this temperature dependence. However, it is not trivial to obtain the temperature correlations of the elastic moduli, in particular in the case of coatings, since they can completely differ from the ones of the corresponding bulk materials. For this reason, here, the obtained CTE refers to the mean value of the CTE over the imposed temperature range.
\newline
Residual stresses can possibly be superposed to $\sigma _{f}$. If, instead of $R(T)$ and $R_{0}$, the curvature radii after and before deposition are considered in eq. \ref{eq1}, the residual stress can be determined.
\newline
The key experimental step is, evidently, the correct determination of the sample curvature; it is performed exploiting an array of parallel laser
beams. The procedure of obtaining the curvature is analyzed in the simplest case of two parallel laser beams, with an initial spacing $D_{0}$, that
impinge on a sufficiently reflective surface at two points A e B, at a nominal incidence angle $\theta $ with respect to the normal to the surface
(see Fig. \ref{Fig4}a). If the surface is flat ($R=\infty $), reflection occurs at an angle $2\theta $ and the two beams are again parallel, at
distance $D_{0}$. If the surface has a convex shape with a finite radius $R$, the two reflected beams are no longer parallel. Simple reflection implies
that the angle $\alpha $ between the normals to the surface at the reflection points A and B is related to the nominal incidence angle $\theta$
as: 
\begin{equation}
\sin \alpha =\frac{D_{0}}{2R\cos \theta }  \label{eq3}
\end{equation}%
and that the angle between the two reflected laser beams is $4\alpha$. The beams are finally detected by a CMOS sensor, supported by a measurement
arm of length $A$. The beams produce on the CMOS screen two spots, at distance d = d$(R)$; for a perfectly flat surface d = d$(R=\infty )$ = d$_{\infty }=D_{0}$. It is intuitive, and it is detailed in the Appendix, that the absolute sensitivity $d$d$/d(1/R)$ increases with the arm length $A$.
However, a larger arm length also implies a stronger sensitivity to vibrations and the need of a larger sensor (although d$_{\infty }$ does not
increase).The insertion of a converging lens, of focal length $F$, in the measurement arm has been considered, with the objectives of reducing the
spot distances, to allow a smaller CMOS sensor, and to limit the arm length $A$, without losing sensitivity. The lens is at distance $L$ from the sample,
and the light sensor is at a further distance $K$ : $A=L+K$ (see Fig. \ref{Fig4}b). In a typical experiment the sample has an initial radius of curvature $R_{i}$, due to the residual stresses, which makes the beams to be reflected with an angle $\alpha _{i}$; the distance between the spots is shifted by $\Delta $d$_{i}$ from d$_{\infty }$. Imposing a temperature variation, the curvature changes to the final value $R_{f}$, the angle changes to $\alpha_{f}$, and the distance between the spots undergoes a further shift $\Delta$d$_{f}$. The change of the curvature radius can be derived from the final displacement $\Delta$d$_{f}$ exploiting the classical matrix optics used in ray tracing algorithms. This method adopts two approximations: the paraxial one, i.e. the smallness of the deviation angle of the beam with respect to the optical axis of the system, and the thin lens one. Both approximations are fully appropriate: firstly, since (see eq. \ref{eq3}) $D_{0}\sim 1$cm and $R$ is at least several meters, $\sin \alpha \precsim 10^{-3}$; secondly, the radius of curvature of the adopted collecting lens is much larger than its thickness. The beams on the sample surface and on the CMOS screen are related by a transfer matrix as follows:
\begin{equation}
\begin{bmatrix}
z^{\prime } \\ 
\alpha ^{\prime }
\end{bmatrix}
=
\begin{bmatrix}
1-K/F & L(1-K/F)+K \\ 
-1/F & 1-L/F
\end{bmatrix}
\begin{bmatrix}
z \\ 
2\alpha 
\end{bmatrix}
\label{eq4}
\end{equation}
where $z$, $\alpha $, $z\prime $ and $\alpha \prime $ are the distance of the beam and its deviation angle, from the optical axis, respectively on the
sample and on the CMOS. Eq. \ref{eq4} gives  
\begin{equation}
z^{\prime }=(1-K/F)z+2\alpha \lbrack L(1-K/F)+K]  
\label{eq5}
\end{equation}
and $\Delta $d$_{f}$ is given by (see Fig. \ref{Fig4}b)  
\begin{equation}
\Delta d_{f}=2 \times (z^{\prime }(2\alpha _{f})-z^{\prime }(2\alpha_{i}))\quad .  
\label{eq6}
\end{equation}
Combining with eq. \ref{eq3} we obtain 
\begin{equation}
\frac{\Delta d_{f}}{D_{0}}B=(\frac{1}{R_{f}}-\frac{1}{R_{i}})  
\label{eq7}
\end{equation}
where $B=(\cos (\theta ))/(2[L(1-K/F)+K])$ is a pure geometrical factor that depends on the
angle of incidence of the beams, on the arm length and on the presence of the focusing lens. If the lens is removed, eq. \ref{eq7} becomes the
standard equation  for measuring the curvature change of a sample by a 2D array of parallel laser beams \cite{Hang2010, Chason2001}: 
\begin{equation}
\frac{\Delta d_{f}}{D_{0}}\frac{\cos \theta }{2A}=(\frac{1}{R_{f}}-\frac{1}{R_{i}})  \label{eq8}
\end{equation}
\begin{figure}[!t]
\centering
\includegraphics[width = 0.6\columnwidth]{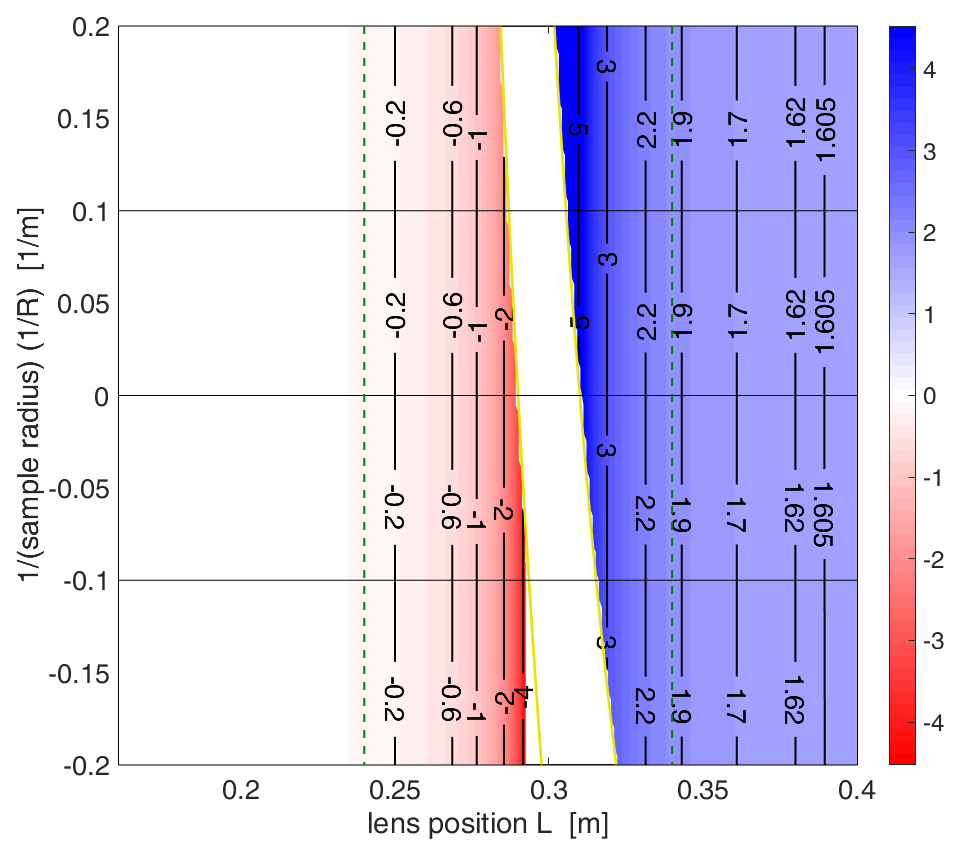}
\caption{Relative sensitivity $d(\Delta d_{f}/$d$_{\infty })/d(1/R)$ [m] map for different lens positions $L$, sample curvature radii $R$, and for fixed arm length $A = 0.4$ m and focal length $F = 0.1$ m. Continuous yellow lines delimit the white region where the measurements are not possible (see Appendix). Dashed green lines indicate the positions $L = 0.24$ m and $L = 0.34$ m of Fig. \ref{Fig_A1}c and Fig. \ref{Fig_A1}b respectively. The right border ($L = 0.4$ m) is the lens-less case of Fig. \ref{Fig_A1}a.}
\label{Fig_x}
\end{figure}
In our experimental setup, $K$,$L$ and $F$ can be varied; an optimization process has been performed, as detailed in the Appendix. Both cases $K<F$
and $K>F$ have been considered. The image on the CMOS sensor can be shrinked, such that the spot distance for a flat specimen, d$(R=\infty )=$ d$_{\infty }$, becomes smaller than $D_{0}$. The absolute sensitivity $d\left(\Delta d_{f}\right) /d(1/R)$ has to be assessed against the physical pixel
size of the sensor; however, the performance of the experimental configuration is better characterized by the relative sensitivity $d(\Delta d_{f}/$d$_{\infty })/d(1/R)$, which has to be assessed against the sensor resolution, in terms of the number of sensor pixels. Fig. \ref{Fig_x} presents the relative sensitivity obtained for a fixed $A=0.4$ m and a fixed $F=0.1$ m, varying $L$ between $0.16$ m and $0.4$ m (the latter distance is the lens-less case). The distance $L=0.34$ m has been identified, which allows an image shrinkage by a factor of more than 2 (as shown in the Appendix), therefore a significantly smaller sensor, with a
relative sensitivity which is larger, by over 20\%, than that of the lens-less case with the same $A$ (as shown by Fig. \ref{Fig_x}). This configuration ($A=0.4$ m, $F=0.1$ m and $L=0.34$ m) is adopted in our measurements; it is suitable up to strong curvatures ($R$ down to 5 m or even less). As it can
be seen from Fig. \ref{Fig_x}, if the curvature is not very strong  (e.g. $R$ above 10 m) the lens can be shifted to slightly smaller values of $L$, obtaining a further boost of the relative sensitivity.
\newline
Operationally, a small array of laser beams is adopted. From the image collected by the CMOS sensor, the positions of the spots due to the various
beams are obtained by standard image analysis procedures, namely the centroid determination, and eq. \ref{eq7} is exploited, taking for the value
of $\Delta d_{f}/D_{0}$ the ratio averaged on all adjacent spots. 
\newline
The obtained curvature radius gives, by eq. \ref{eq1}, the total stress in the sample, from which, by eq. \ref{eqCTE}, the average $CTE_{f}$ over the
imposed temperature range can be obtained. It must be remembered that the measured CTE refers to the $in-plane$ component of the liner expansion
thermal coefficient. In anisotropic sample the in-plane component can significantly differ from the out-of plane CTE, that must be determined by
other techniques.
\newline
As discussed in the following section and in Appendix B, different noise sources related to the experimental setup severely affect measurement accuracy. However, also surface roughness can result in inaccuracies of CTE determination when, instead of the substrate uncoated surface, the film surface is probed. Roughness can induce imperfections in the reflected beam spots shape, eventually affecting the accuracy of the spot centroid calculation. In our case, PLD coatings on flat silicon substrates show a very low surface roughness (i.e. few nanometers) that, added to multiple frames average, limit this error source. Thickness inhomogeneities, instead, generally influence measurement accuracy, resulting in an apparent increase or decrease of curvature. All the W coatings analyzed in this work are characterized by a high planarity, of the order of $\pm$ 10\% the mean film thickness. This means that, for 400 nm thick coatings, a variation of $\pm$ 40 nm can eventually result in a change of curvature radius of $\pm$ 400 m. This value is an order of magnitude higher than the commonly measured curvature radii (i.e. few tens of meters), thus introducing a small error in the CTE computation. Thickness inhomogeneities, instead, become critical for micrometric thick coatings.

\section{EXPERIMENTAL SETUP}

\label{Experimental techniques}

\begin{figure}[!t]
\centering
\includegraphics[width = 0.7\columnwidth]{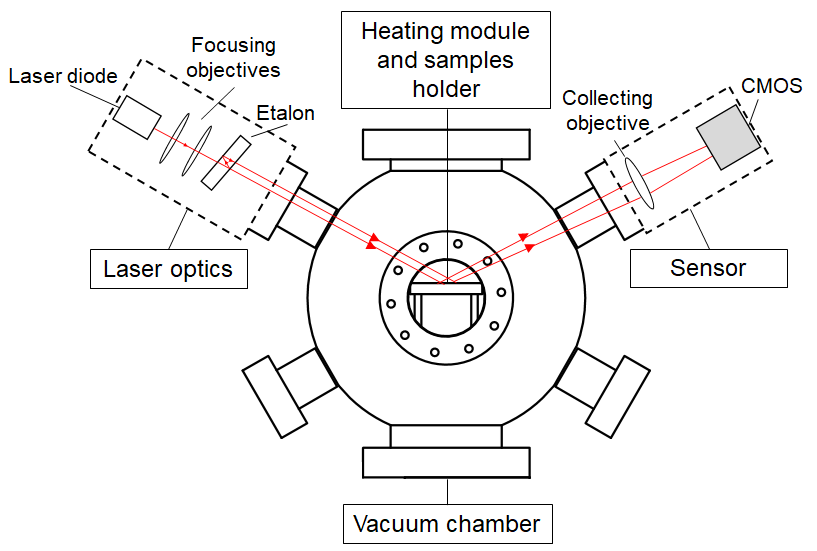}
\caption{Schematic diagram of the apparatus for CTE measurements. The heating module is schematically shown in the inset of figure \ref{Fig3}.}
\label{Fig1}
\end{figure}

The schematic diagram of the apparatus developed for CTE measurement is shown in Fig.\ref{Fig1}. It consists of three main parts: a set of laser optics, a vacuum chamber for thermal annealing processes and a sensor for laser beams positions measurement. 

\subsection{Laser beams array generation and collection systems}

The laser beam array is generated by coupling a laser diode ($\sim$ 5 mW output, 630 nm wavelength) and a pair of etalons. The first etalon multiplies the input laser beam in a direction, while the second etalon, oriented at 45$^\circ$ with respect to the first one, duplicates the 1D array in the other direction, obtaining a 2D parallel laser beams array. 
 In our case, we create a 2 x 2 array of 1 cm equally spaced laser beams for a total coverage of 1 x 1 cm$^2$ measurement area. The array strikes with an angle of incidence of 60$^\circ$ at the center of the substrate polished surface. The measurement position is kept constant during the entire analysis process. The reflected beams are recorded by the CMOS camera, through a collecting lens, as discussed above. The adopted camera is characterized by a 4/3, 1.3 Megapixel sensor with a 1024 x 1248 digitized image. The acquisition rate of the sensor is 10 fps in the full format, but it can be further increased up to 200 fps if only certain regions of interest are selected. In this way, multiple measurements for a certain temperature step can be acquired, so that the signal can be averaged on successive frames in order to reduce the overall noise. The position of the beams on the CMOS is followed by the determination of the centroid of each laser spot. The centroids are determined by a classical centroid of intensity algorithm, which weights the intensity of each pixel in the irradiation area over the total irradiated pixels. It has to be noted that the accuracy of the method is deeply affected by noise sources (see Appendix B), such as vibrations from the vacuum system or gas flow and fluctuations of pixels intensity. The use of multiple laser beams array is thus crucial to guarantee high measurement accuracy. With respect to standard laser scanning systems, where a single laser beam is rastered across the entire sample surface \cite{Flinn1987}, the use of multiple beam array results in measurements that do not depend on the absolute position of each laser spot on the sensor \cite{Floro1996}. The laser beams strike all at the same time on the sample surface and the differential beam spacing between adjacent spots, which is less sensitive to the sample vibrations than the absolute position of the beam, is adopted to measure the change of sample curvature. With our setup, we obtain an accuracy of the beam spacing measurement of 0.09 pixels, that with our sensor of 6.66 x 6.66 $\mu$m/pixel stands for $\pm$0.5 $\mu$m maximum deviation (see Appendix B).


\subsection{Vacuum chamber and heating module}
\begin{figure}[!t]
\centering
\includegraphics[width = 0.5\columnwidth]{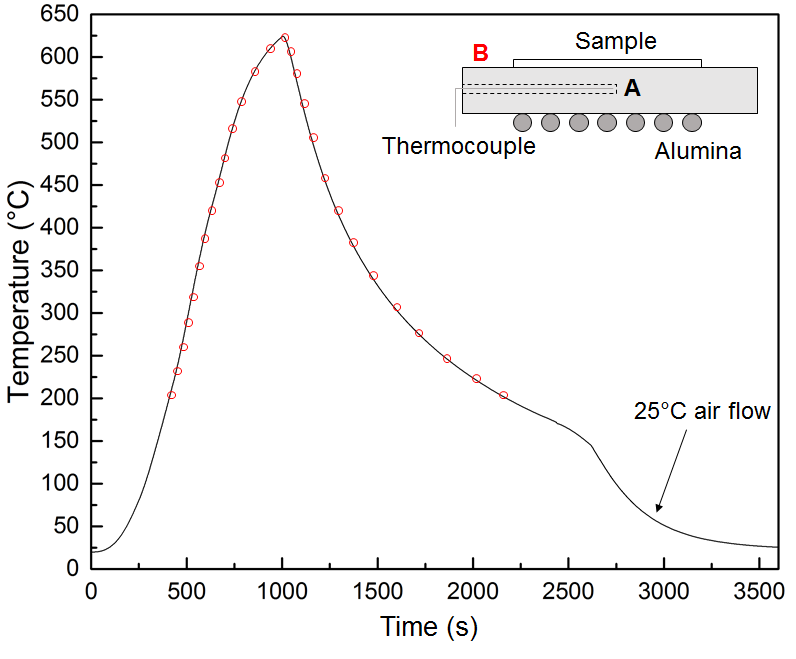}
\caption{Temperature measurement of the heating module during the maximum temperature ramp adopted for this work. The straight line refers to the measurement by the K-type thermocouple in the point A; the circle scatter points indicate the temperature measured by the pyrometer on the top of the heating plate (point B)}
\label{Fig3}
\end{figure}
The vacuum chamber equipped with the laser beams array generation and collecting systems described above accommodates the heating module for controlled thermal annealing processes (see Fig. \ref{Fig1}). The vacuum chamber is a 300 mm spherical chamber supplied with two 2” viewport flanges at 60$^\circ$ orientation for the input and the output of the laser beams. Other two symmetrical flanges provide connections with the vacuum system and the gas inlet for controlled atmosphere treatments. A coupled rotary and turbomolecular pumps are exploited to guarantee a base pressure of 5 x 10$^{-6}$ mbar during each thermal treatments. The heating module can reach up 1200 $^\circ$C. The temperature is measured by a thermocouple (type K) place under the sample in the middle of the holding plate (5 mm thick). A standard temperature ramp is shown in Fig. \ref{Fig3}. The uniformity of temperature along the plate thickness and lateral dimension has been assessed by a pyrometer measurement (red circle marks). The heating and cooling rate are fixed at 40 $^\circ$C/min and 20 $^\circ$C/min respectively. Temperature is measured every 0.5 s by an external acquisition system triggered with the CMOS data acquisition by an ad-hoc developed Labview interface. In this way, the centroid positions of each spot are automatically synchronized with temperature data.

\begin{table*}[t]
\centering
{\relsize{-2}
\begin{tabular}{lccccccc}
\toprule
Structure & Morphology & Deposition conditions & Composition & D (nm) & $\rho$  (g cm$^{-3}$) & Thickness ($\mu$m) & $M$ (GPa)\\
\toprule
n-W       & Compact & Vacuum       &         metallic W                    &      16      &      18      & 0.4      & 527\\
u-n-W	& Compact & Vacuum       &  W 90$\%$ - Ta 10$\%$  &      11	     &      13      & 0.4      &  500    \\
u-n-W	& Compact & He 70Pa annealed  &  metallic W                 &        7	     &      12      & 0.38    &  353     \\
a-W	      &  Compact &He 70Pa       & metallic W	                         &        $<$ 2 nm      &	11     & 0.41     &   227 \\
a-W      &  Compact & $O_2$ 5Pa   &  W-O                             &       $<$ 2 nm    &      11	   & 1         &  265  \\
a-W	     &   Porous & He 100Pa      & metallic W	                         &   $<$ 2 nm      &       9	 & 0.43      & 189  \\
\bottomrule
\end{tabular}
}
\caption{Samples investigated in this work. Biaxial modulus is derived from Brillouin spectroscopy \cite{Besozzi2016}, coating thickness by SEM analysis, crystallites dimension $D$ by XRD \cite{Besozzi2016, DellasegaJAP} and mass density $\rho$. The coatings are deposited by PLD in vacuum or in presence of background gases (He, O$_2$) at different partial pressures as reported in \cite{Pezzoli2015, DellasegaJAP}.}
\label{Tab2}
\end{table*}

\section{RESULTS AND DISCUSSION}

Firstly, we tested the performances of our experimental setup by investigating the CTE of different coating materials. In particular, we analyze thermally evaporated silver (Ag) films as the ones investigated in \cite{Lima1999}. The Ag coatings have been deposited on a Si(100) 500 $\mu$m thick double side polished substrate. The thicknesses of the coatings have been determined by Scanning Electron Microscopy (SEM), being all sub-micrometric. Since Ag films have been deposited in very similar conditions of \cite{Lima1999}, the biaxial modulus of the deposited material have been chosen between 50 $\pm$ 10 GPa. The CTE of Ag films was investigated in the 25 - 150 $^\circ$C temperature range. We obtained a CTE of 38 $\pm$ 4 10$^{-6}$ K$^{-1}$, that well fits the result obtained in literature of 33 $\pm$ 4 10$^{-6}$ K$^{-1}$. Starting from this result, we proceeded with the characterization of W based coatings as described below.

\subsection{Samples preparation, structural and elastic properties}

\begin{figure*}[!t]
\centering
\includegraphics[width = 0.8\columnwidth]{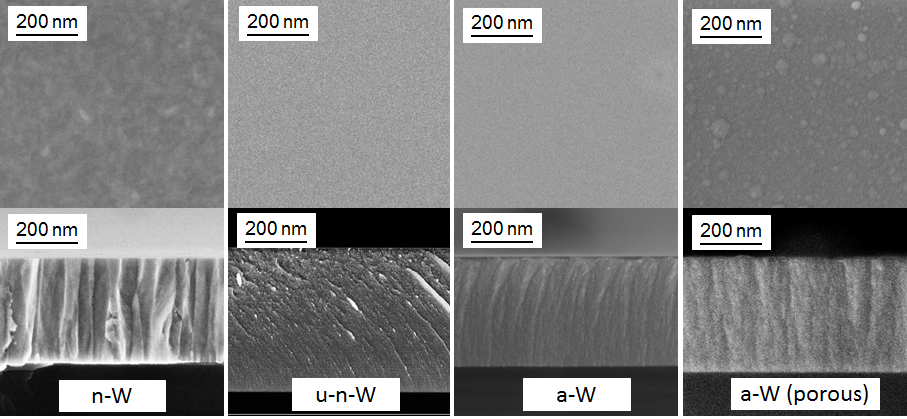}
\caption{From the left: SEM top view and cross section images of nanocrystalline columnar W, ultra-nano-crystalline W, compact amorphous-like W and porous amorphous-like W samples analyzed in this work.}
\label{FigSEM}
\end{figure*}

All the nanostructured W coatings were deposited by PLD on silicon (Si) (100) substrates 500 $\mu$m thick. For the $\sigma_f$ computation of eq. \ref{eq1} we consider $E$ = 160 GPa, $\nu$ = 0.28 and CTE = 2.7 10$^{-6}$ K$^{-1}$ for this type of substrate \cite{Okada1984}. The tailored nanostructure is obtained tuning the background gas pressure (i.e. helium (He) and oxygen (O$_2$)) during deposition. For more details about the deposition process of these coatings refer to \cite{Besozzi2016, Pezzoli2015, DellasegaJAP}. As summarized in Tab. \ref{Tab2}, the change of film nanostructure from nanocrystalline (n-W) to amorphous-like (a-W) is achieved increasing the background gas pressure from vacuum conditions to 100 Pa of He pressure. While up to 75 Pa of He the a-W morphology is still compact, at 100 Pa of He pressure the a-W starts to become porous. O$_2$ is exploited to obtain again the a-W structure. As reported in \cite{Pezzoli2015}, at O$_2$ pressures of 5 Pa the W/O ratio is sub-stoichiometric (i.e. about 2.4) and the film preserves its metallic nature. As comfirmed by X-ray Diffraction (XRD) analysis, the W-O sample analyzed in this work is characterized by an amorphous-like structure. Finally, the ultra-nano-crystalline (u-n-W) structure is formed by thermal annealing of a-W pure W coatings over their recrystallization temperature (i.e. 650 $^\circ$C) or by adding tantalum (Ta) as solid solution during coatings deposition. The detailed study of the recrystallization behavior of a-W and the effect of Ta alloying on coating structure is reported in \cite{Besozzi2016}. Here, we limit our study to a 650 $^\circ$C thermally annealed a-W and an u-n-W coating with 10$\%$ of Ta concentration. The morphology and structural evolutions of the coatings when going from n-W to compact and porous a-W are highlighted by Scanning Electron Microscopy (SEM) analysis summarized in Fig. \ref{FigSEM}. SEM analysis were exploited also to determine the coatings thicknesses that are summarized in Tab \ref{Tab2}. 
\newline
These three different nanostructures are characterized by different mean crystallite sizes ($D$), determined by XRD analysis using the Scherrer correlation. As shown in Tab. \ref{Tab2}, $D$ goes from 16 nm in the case of n-W to below 2 nm for a-W, assuming values between 7 and 11 nm in u-n-W samples. The film mass density $\rho$ of pure nanostructured W coatings was determined by quartz microbalance measurements during deposition \cite{DellasegaJAP}. $\rho$ decreases from 18 g cm$^{-3}$ to 9 g cm$^{-3}$ when going from n-W to a-W. On the contrary, the mass densities of W-Ta and the u-n-W annealed samples were derived from the lever rule and numerical simulations respectively, as described in \cite{Besozzi2016}. No direct measurmente of the $\rho$ for the W-O coating is available. However, since its nanostructure and morphology were found to be similar to the one of pure a-W, $\rho$ is fixed at the same value of 11 g cm$^{-3}$. 
\newline
The elastic properties (i.e. the biaxial modulus $M = E/(1-\nu)$) of each W coating have been determined by Brillouin spectroscopy. For a detailed description of the derivation method see \cite{Besozzi2016}. As it can be seen in Tab. \ref{Tab2}, $M$ is strictly related to the changes of film mass density and crystallites dimension, going from 527 GPa for n-W to 189 GPa for a-W. It has been shown \cite{Besozzi2016} that in the regions where the mass density does not significantly change, $D$ is the key parameter that affects the elastic behavior of the material and vice versa. This explain why different a-W and u-n-W samples show different values of $M$. 

\subsection{Residual stress of nanostructured W coatings}

\begin{figure}[!t]
\centering
\includegraphics[width = 0.6\columnwidth]{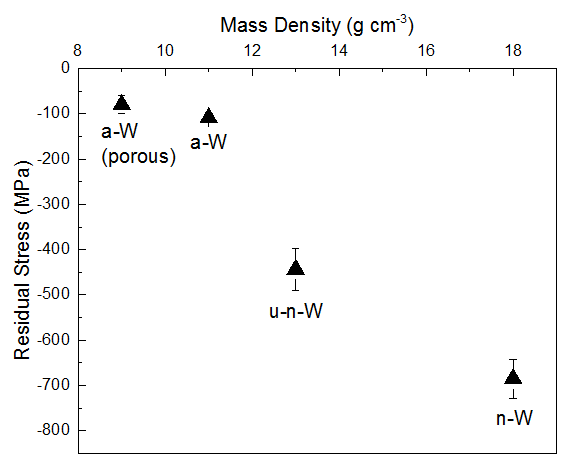}
\caption{Residual stresses of nanostructured W coatings plotted versus film mass density. The lower density of the film, which is related to a lower energy of the ablated particles, implies a lower state of compressive stress.}
\label{FigRes}
\end{figure}

The initial state of stress of the coatings is obtained by measuring the curvature change between the uncoated and coated Si wafer at room temperature. In the case of n-W, for instance, we found an initial state of compressive residual stress of 684 $\pm$ 42 MPa. The compressive stress is in agreement with the residual stresses found in other coatings deposited by PLD. As pointed out in different works \cite{Bonelli2000, Teixeira2002, Ganne2002, Lackner2004, Cibert2008}, the higher the energy of the ablated particles, the higher the compressive residual stresses are found in the PLD coating. For this reason, it is clear that columnar nanocrystalline W samples, that are deposited in vacuum conditions, are characterized by a higher compressive residual stress than the amorphous ones. In the case of a-W coatings, indeed, the presence of a background gas during deposition implies a loss of the ablated particles energy before impinging on the substrate. As a result, the particles on the substrate are not enough energetic to be as closely packed as in columnar film; the coatings grow with completely different structures and morphologies, and they are characterized by a lower state of stress. We thus evaluate the residual stresses of our PLD W coatings. Since the stress is strictly related to the coating thickness, we reported in Fig. \ref{FigRes} only the residual stresses of the coatings with aproximately the same thickness (i.e. 400 nm). They are plotted versus film mass density. As it can be seen, the trend observed is fully consistent with the explanation proposed herein. The residual stress, indeed, drops from 684 MPa for n-W, where the highest mass density is observed, to around 80 MPa for the porous a-W structure when the mass density becomes the 50$\%$ the n-W one. 

\subsection{CTE characterization of nanostructured W coatings}
\begin{figure}[!t]
\centering
\includegraphics[width = 0.7\columnwidth]{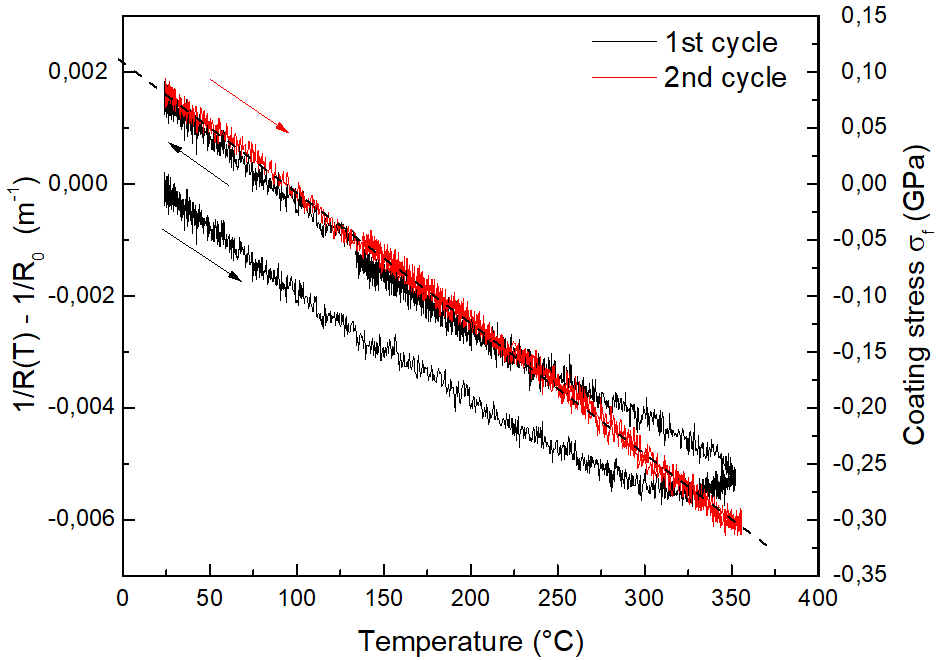}
\caption{Thermal cycles performed on a-W sample. The black and the red lines are the first and the second thermal cycles respectively. The blue dashed line represents the linear interpolation of the stress-temperature curve adopted to derive the mean CTE of the coating (see eq. \ref{eqboh}).}
\label{Fig5}
\end{figure}
\begin{figure}[!t]
\centering
\includegraphics[width = 0.5\columnwidth]{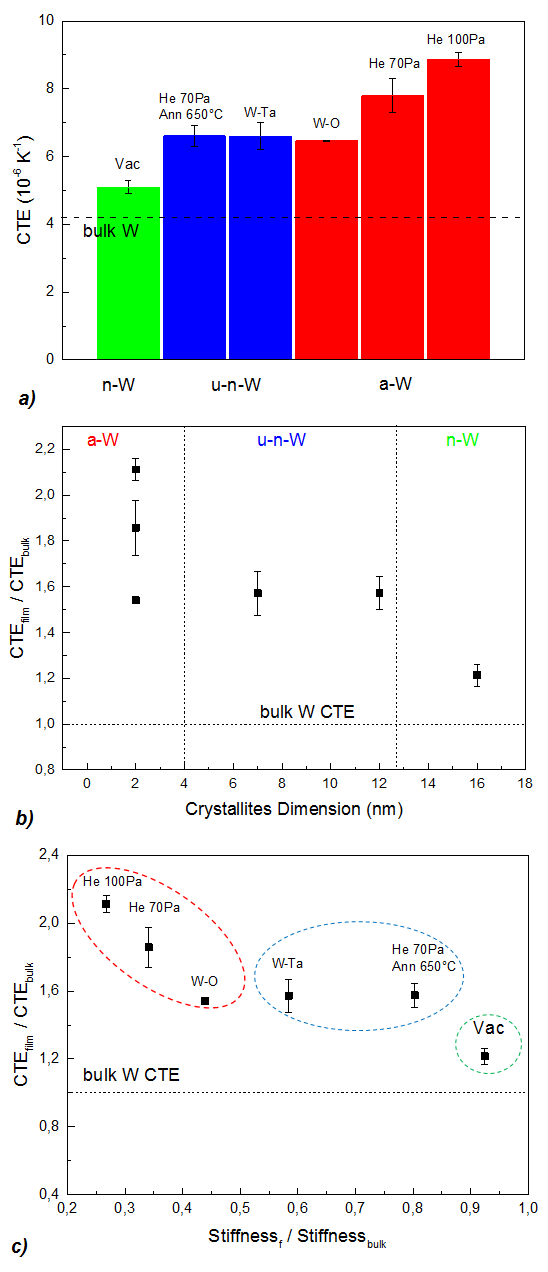}
\caption{a) CTE of the analyzed samples. b) Dependence of he CTE to the crystallite dimension of the sample c) The ratio between the measured and the bulk CTE plotted versus the ratio of the coating and bulk stiffness. The three regions of n-W, u-n-W and a-W are highlighted by the green, the blue and the red dotted lines respectively.}
\label{Fig6}
\end{figure}

For the CTE characterization all the elastic properties for both the film and the substrate are considered temperature independent. In this way only a mean value of the CTE over the imposed temperature range can be determined by this method. The maximum temperatures reached for each samples have been chosen depending on the recrystallization temperatures of each phase. It has been shown \cite{Besozzi2016, Pezzoli2015} that the crystallization process of a-W starts even at 450 $^\circ$C, which is well below the bulk W recrystallization temperature (i.e. 1400 $^\circ$C). Therefore, a-W coatings are annealed below 400 $^\circ$C in order to avoid the formation of the $\alpha$-W phase, hindering crystallites growth that triggers the formation of the u-n-W structure. For n-W and u-n-W no phase changes are observed below 1000 $^\circ$C. However, the choice of the maximum annealing temperature is also strictly dependent by the type of the substrate material. In the case of Si, it is known that above 650 $^\circ$C tungsten silicide can form at W-Si interface \cite{Tsaur1984}, deeply affecting the CTE measurements. For this reason all the measurements for n-W and u-n-W samples are limited to a maximum temperature value of 650 $^\circ$C. 
\newline
As an example, Fig. \ref{Fig5} shows the thermal stress cycles measured for a-W sample. The sample is heated from room temperature to about 360 $^\circ$C and then cooled down again to room temperature. This cycle has been performed twice. The thermal stress is plotted versus annealing temperature. As it can be seen, the negative intensity of the thermal stress stands for a compressive state of stress in the coating. This is always the case when the film shows a higher CTE than the one of the substrate material. In this way, the coating tries to dilatate but it is constrained by the substrate, developing a compressive stress which grows as the annealing temperature increases. On the contrary, the stress becomes tensile during the cooling cycle of the sample. A clear difference can be found between the first and the second thermal cycle around the maximum annealing temperature. During the first heating cycle the compressive thermal stress grows linearly with temperature till around 290 $^\circ$C. Over 290 $^\circ$C a clear nonlinear behavior is observed. This trend is exhausted during cooling, when the tensile state of stress starts to grow again linearly with cooling temperature. This characteristic feature can be explained by the beginning of stress relaxation processes, which lead to plastic deformation in the 290 - 360 $^\circ$C temperature range. The relaxation processes could indicate the origin of grain growth or the triggering of defects diffusion processes which can be present in a-W even at very low temperatures. The relaxation process is driven by an enhanced surface and bulk atoms diffusion, which continues till the atoms reach their equilibrium positions. The consequent volume shrinkage associated with the developed plastic flow results in the development of a tensile state of stress which is highlighted in the stress-temperature curve in Fig. \ref{Fig5} by the deviation from linearity during heating. After plastic deformation behavior takes place, the sample is not able to recover the same state of stress during cooling. Due to the development of relaxation irreversible changes of the layer structure, the non linear part of the stress curve can not be used to derive the CTE of the material. This trend is not observed when the sample is subjected to a second thermal cycle between the same temperatures, clearly indicating that no more relaxation processes take place. In this way, the total stress temperature curve (i.e. heating and cooling) can be fitted by a linear regression in order to derive the contribution $d\sigma /dT$ of eq. \ref{eqCTE} (dotted blue line). Once the slope is determined, using the biaxial modulus of the film summarized in Tab. \ref{Tab2}, the mean CTE of the coating are obtained. The behavior of the thermal stress upon heating is found in all the analyzed coatings. However, we observe that for higher increasing initial residual stress, higher number of cycles are needed to exhaust the relaxation process of the stress and to obtain the linear trend of stress versus temperature during both heating and cooling.  
\newline
The results are shown in Fig. \ref{Fig6}a. All the CTE of the coatings lie above the bulk value of 4.2 x 10$^{-6}$ K$^{-1}$ reported in literature for polycrystalline W \cite{Nix}. A clear dependence of the CTE by the nanostructure is found. n-W has a mean CTE of 5.1 x 10$^{-6}$ K$^{-1}$, which is close to the bulk one. u-n-W samples are characterized by an increase of the CTE to around 6.6 x 10$^{-6}$ K$^{-1}$. Finally, for a-W samples, CTE reach a maximum value of 8.9 x 10$^{-6}$ K$^{-1}$, which is almost twice the bulk one. The mean value and the error bars associated to each point are evaluated by the multiple measurements performed on each sample. The uncertainty related to the CTE derivation from the geometric values via Stoney's equation (see eq. \ref{eq1}) obviously also depends on the uncertainties related to the elastic moduli and the thicknesses which must be independently measured. 
\newline
Nanocrystalline metals are used to show a higher CTE with respect to the one of the crystalline counterpart \cite{Lima1999, Lu1995, Marques2003}. With respect to a crystalline bulk W, the presence of a higher fraction of interfaces between the small grains deeply affect the properties of the material \cite{Daniel}. The weaker bonding of grain boundaries atoms modify the interatomic potential, lowering it and making it more asymmetrical. The net result is a favoured movement of the atoms around their lattice positions upon heating. This means an enhancement of the CTE, which is thus strictly related to the volume fraction of grain boundaries. It has been shown that, in the case of nanocrystalline metallic films, the CTE at grain boundaries can even increase 2 - 5 times the crystalline value \cite{Birringer1988, Klam1987}. This dependence of the CTE with the crystallites dimension is shown in Fig. \ref{Fig6}b.      
As it can be seen, the overall observed behavior is that as $D$ decreses the CTE grows. n-W sample shows a CTE 1.2 times the bulk one, increasing up to 2.1 times for a-W samples. This trend is qualitatively and quantitatively in accordance with the reported dependence by $D$ of the CTE of some metallic films investigated in literature. As reported in \cite{Lu1995}, for example, copper films with 8 nm grains show a 1.8 times higher CTE than the corresponding monocrystalline structure. In our case, u-n-W coatings with $D$ between 7 and 11 nm are characterized by a CTE 1.6 $\pm$ 0.1 times higher the bulk W one. 
However, when the amorphous regime is reached, the dependence of the CTE by the grain boundaries fraction becomes not consistent to explain its further increase up to 2.1 times the crystalline value. The monotonically increasing behaviour of the CTE in the a-W region is not worthy, since the investigation of the CTE of amorphous materials still leads to controversial results in literature. In some cases \cite{Lima1999, Magisa2014, Tong1992}, starting from the coarse grained structure, an increase of the CTE is observed as $D$ decreases, but, when the amorphous region is reached, a drop of the CTE is obtained. On the other hand, other works \cite{Lu1995, Daniel, Hay2006, Miller2010}, in accordance with the trend observed in this work, reported a still higher CTE of the amorphous phase wih respect to the nanocrystalline one. This is consistent with higher
mean interatomic distance, which means a lower binding energy. On the other hand, the mean interatomic potential can be affected by the density of defects, that in turn are related to the tensile or compressive residual stress \cite{Zoo2006, Chaplot}. The porosity of the film can be a key parameter in driving the thermal expansion of the coating, inducing preferred dilatation directions, with a net result of an increase of material CTE \cite{Miller2010}. 
In our case, for a-W samples, in particular for the a-W samples deposited at 70 and 100 Pa of He, the higher porosity degree of the films with respect to the other samples can be identified as the main parameter to properly justify the increase of the CTE above the u-n-W values. This increasing porosity is remarked by the measured drop of the film mass density, that goes from 18 g cm$^{-3}$ in n-W to 9 g cm$^{-3}$ in the amorphous region. The CTE trend we observed for W in the different n-W, u-n-W and a-W is consistent with the behavior of other bcc metals reported in literature, such as chromium and tantalum \cite{Knepper2007, Daniel}.
\newline
Finally, as the variations of $D$ and $\rho$ modify the thermal expansion properties of the material, they also affect the elastic behavior of the samples \cite{Besozzi2016}. Here, referring to Fig. \ref{Fig6}c, we want to highlight the relationship between the CTE and the stiffness of nanostructured coatings. So, in Fig. \ref{Fig6}c the CTE is plotted versus the film to bulk stiffness ratio (i.e. E$_{film}$/E$_{bulk}$). Again, the three n-W, u-n-W and a-W regions are clearly distinguishable. n-W sample is characterized by around 92$\%$ of bulk stiffness; in the u-n-W region the stiffness ratio goes from 80$\%$ to 60$\%$, while for a-W it goes below 47$\%$, down to 27$\%$.  As a general qualitative trend, the softer the material, the higher the CTE. This well known stiffness-CTE behaviour is reported in several literature works \cite{Zoo2006, Lima1999, Marques2003}. However, this relationship is not linear as it can be expected from eq. \ref{eq2}. The deviation from the linear proportionality can be thus attributed again to the interplay between the crystallites dimension and the mass density of the material. 

\section{CONCLUSIONS} 

In this work we performed a systematic study of the residual stresses and the coefficient of thermal expansion of nanostructured W based coatings deposited by PLD, with the aim of elucidating the correlation between the CTE, the residual stresses, the structural (i.e. crystallites dimension, mass density) and elastic properties of the materials. In particular we analyzed pure W, W-tantalum and W-O coatings with different nanostructures. In order to obtain the residual stress and the CTE of the coatings, we developed a novel experimental setup based on the thermally induced substrate curvature method. All the W coatings deposited by PLD are characterized by a compressive residual state of stress. The stress is strictly correlated to the specific nanostructure, becoming lower as going from n-W to a-W, due to a lower energy of the ablated particles. We found that all the analyzed samples show a higher CTE than the corresponding bulk form. n-W shows a CTE of 5.1 10$^{-6}$ K$^{-1}$, u-n-W a CTE of 6.6 10$^{-6}$ K$^{-1}$, while a-W a CTE between 6.6 10$^{-6}$ K$^{-1}$ and 8.9 10$^{-6}$ K$^{-1}$. The CTE is thus deeply affected by the crystallites size, growing as the crystallites dimension decreases, where a higher fraction of grain boundaries is present. This trend is fully consistent with the behavior observed for other bcc metals, such as chromium and tantalum. In the amorphous region, where $D$ does not substantially change, the CTE further increases. Here, we highlight the relationship between the CTE and the film mass density. The higher porosity degree, that characterize the amorphous coatings, plays a pivotal role in giving preferential dilatation directions, favouring thermal expansion. In addition, in accordance with literature, we observed that as the material becomes softer, the CTE of the coating increases.

\section*{Acknowledgments}
This work has been carried out within the framework of the EUROfusion Consortium and has received funding from the Euratom research and training programme 2014-2018 under grant agreement No 633053, and from the MISE-ENEA `\textit{Accordo di Programma}' (AdP), PAR2015 and PAR2016. The views and opinions expressed herein do not necessarily reflect those of the European Commission. The research leading to these results has also received funding from the European Research Council Consolidator Grant ENSURE (ERC-2014-CoG No. 647554).

\bibliographystyle{model1a-num-names}
\bibliography{<your-bib-database>}

\section*{Appendix A: optimization of the experimental setup}

The analysis of the measurement configuration, under the paraxial
approximation discussed in the text, is simple in the case in which the
measurement arm, of length $A$, does not include a lens. With reference to
Fig 1b, in this case the distance of the two spots, on the light sensor, is 
\begin{equation*}
\text{d}=D_{0}+2A\sin 2\alpha 
\end{equation*}
i.e., exploiting eq. \ref{eq3}
\begin{equation*}
\text{d}=D_{0}+4A\frac{D_{0}}{2R\cos \theta }\quad .
\end{equation*}
Therefore 
\begin{equation*}
\frac{\text{d}}{D_{0}}=\left( 1+\frac{4A}{2R\cos \theta }\right) 
\end{equation*}
and the relative sensitivity is 
\begin{equation}
\frac{d\left( \text{d}/D_{0}\right) }{d\left( 1/R\right) }=\frac{4A}{2\cos
\theta }\quad .
\label{eqA1}
\end{equation}

Since in the lens-less case d$_{\infty }=D_{0}$, the relative sensitivity of
eq. coincides with the more general definition $d\left( \text{d}/\text{d}_{\infty
}\right) /d\left( 1/R\right) $. Eq. \ref{eqA1} shows that in the lens-less case the
relative sensitivity is simply proportional to the arm length $A$, and
increases when the incidence angle increases. However, for incidence angles
approaching $90^\circ$ (grazing incidence) the measurement becomes very delicate,
and more sensitive to various causes of error. In our set up $\theta =60^\circ$, a
good compromise between sensitivity and robustness of the measurement. The
relative sensitivity is thus simply $4A$, i.e. $1.6$\thinspace m for $A=0.4$ m and $3.2$\thinspace m for $A=0.8$ m. 
\newline
It can be noted that the relative sensitivity does not depend on $D_{0}$.
The resolution of the light sensor is measured by its number of pixels; the
relative sensitivity can be translated into a resolution in terms of $1/R$ (see Appendix B).
The sensor size must obviously be larger than $D_{0}$, but not too much
larger. A full exploitation of the sensor area is achieved when the sensor
size is, say, $1.5\,D_{0}$ to $2\,D_{0}$.The independence of the relative
sensitivity from $D_{0}$ might suggest that the value of $D_{0}$ be
irrelevant. If sensors of different sizes were available, with the same
number of pixels, measurements with different values of $D_{0}$ would have
the same relative sensitivity, i.e. would be of the same quality, provided
the appropriate sensor size was always selected. The only consequence of a
larger, or smaller, $D_{0}$ would be the need, or not, of a larger sample.
This is not completely true. The beams have a finite lateral size, which is
independent from $D_{0}$, and if their distance becomes too small they
cannot any longer be resolved, and the measurement becomes impossible. The
beams become too close when $D_{0}$ is small and/or the sample has a strong
concavity. If the minimum distance at which the beams can be resolved is,
say, 2 mm, when $D_{0}$ is 5 mm the threshold is reached when d$/$d$_{\infty
}$ is reduced to 0.4, while when $D_{0}$ is 10 mm measurements are possible
until d$/$d$_{\infty }$ is reduced to 0.2. The size $D_{0}$=10 mm is adopted
in our set up and in the following analyses. 

\begin{figure}[!t]
\centering
\includegraphics[width = 0.5\columnwidth]{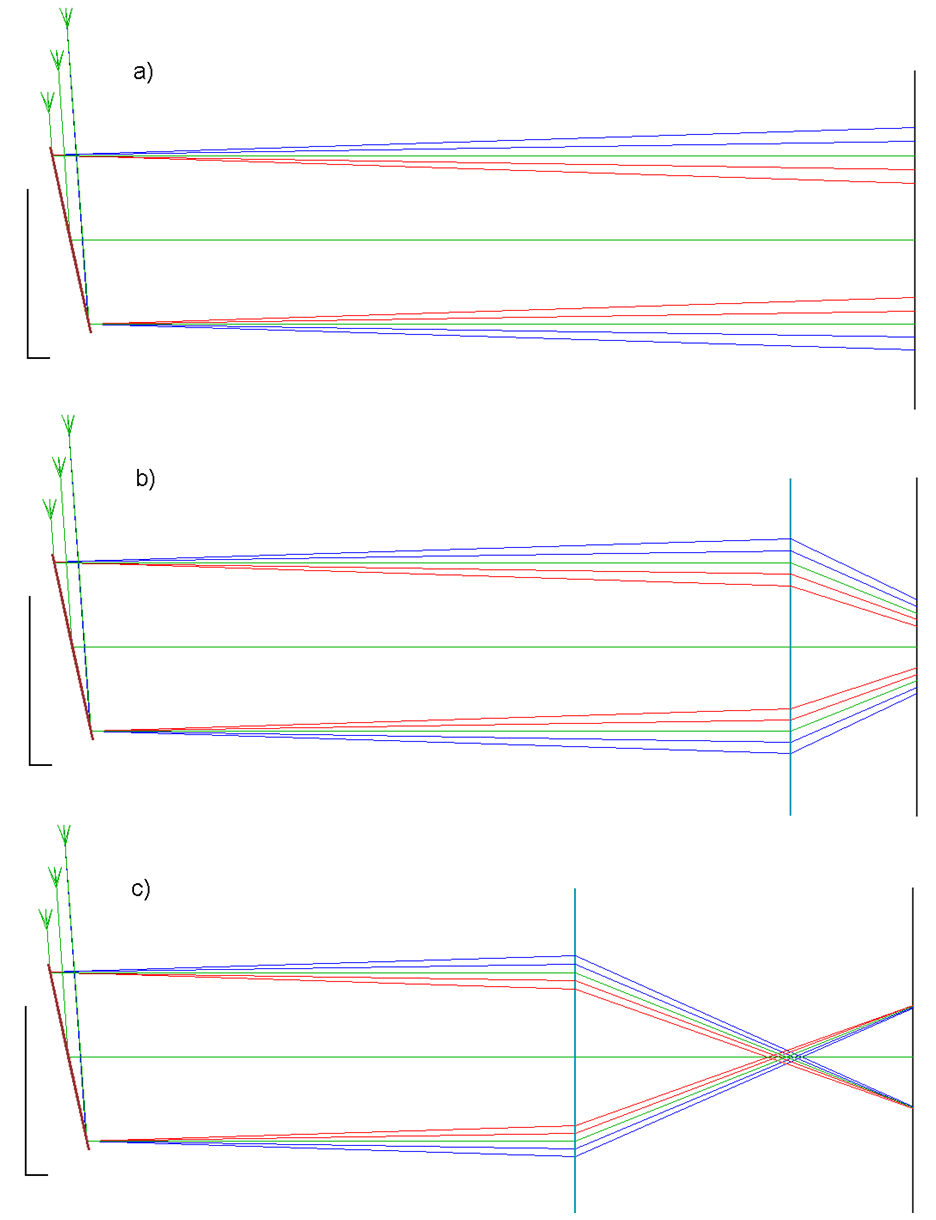}
\caption{Rays trajectories for fixed arm length $A = 0.4$ m, focal length $F = 0.1$ m and beams distance $D_0$ = 10 mm, for the lens-less case (a), ad the lens at $L = 0.34$ m (b) and $L = 0.24$ m (c).  of Fig. \ref{Fig_A1}c and Fig. \ref{Fig_A1}b respectively. In order to improve the readability, the representations are stretched in the vertical direciton (the unit lengths of the two directions are indicated by the black segments).}
\label{Fig_A1}
\end{figure}

\begin{figure}[!t]
\centering
\includegraphics[width = 0.5\columnwidth]{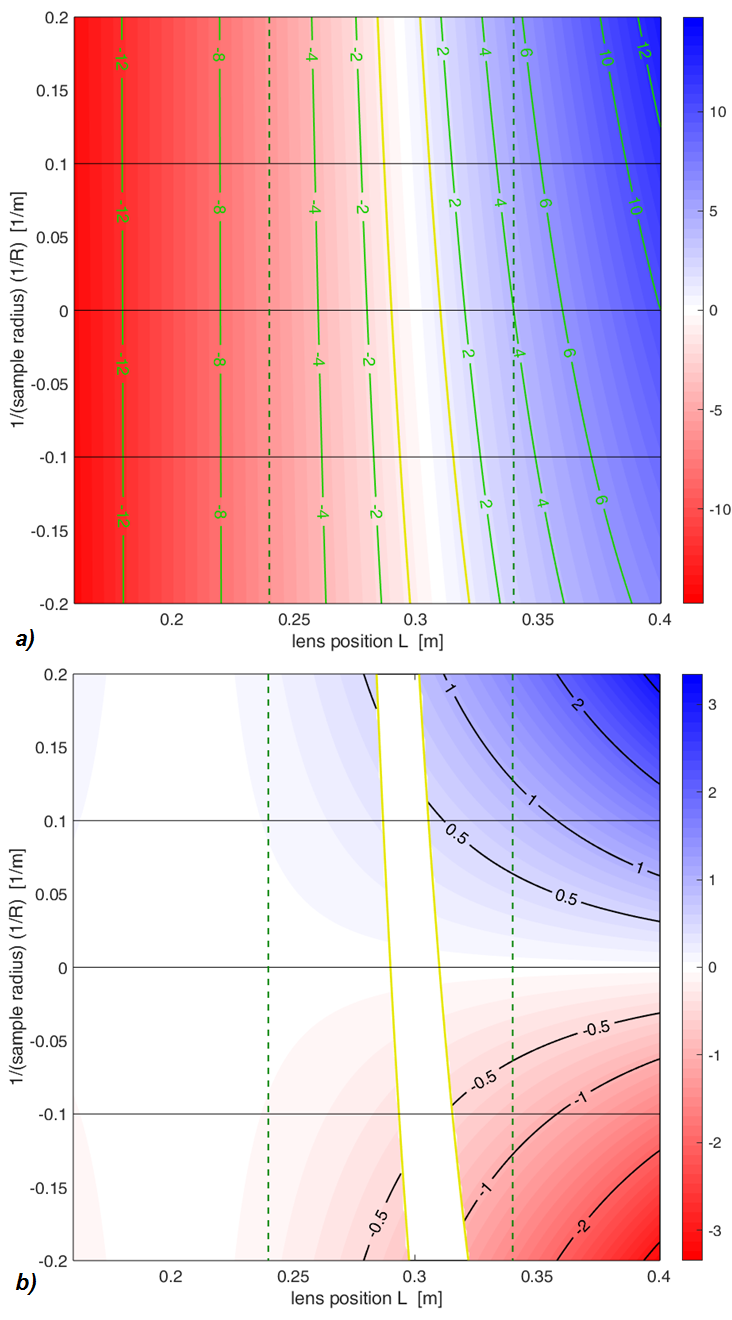}
\caption{Maps of spot distance d ([mm], a)) and spot displacement d - d$_\infty$ ([mm], b)) on the light sensor, for different lens positions $L$, sample curvature radii $R$, and for fixed arm length $A = 0.4$ m and focal length $F = 0.1$ m. Continuous yellow lines delimit the white region where the measurements are not possible because the spots are too close to be resolved. Dashed green lines indicate the positions $L = 0.24$ m and $L = 0.34$ m of Fig. \ref{Fig_A1}c and Fig. \ref{Fig_A1}b respectively. The right border ($L = 0.4$ m) is the lens-less case of Fig. \ref{Fig_A1}a.}
\label{Fig_A2}
\end{figure}

The configuration with the lens is analyzed by an in-house developed Matlab
code which implements the ray tracing technique, under the same paraxial and
thin lens approximations, already mentioned. The results are presented here
in detail  for fixed $A=0.4$ m, $F=0.1$ m and $D_{0}$=10 mm, and varying $L$
between $0.16$ m and $0.4$ m (the latter distance is the lens-less case).
The direct outcomes of the ray tracing analysis are presented in Fig. \ref{Fig_A1}, for the representative cases of $L=0.4$ m (the lens-less case), $L=0.34$ m ($K<F$) and $L=0.24$ m ($K>F$). Some qualitative considerations are
already possible from these figures.
A quantitative analysis is presented in Figs. \ref{Fig_A2} and \ref{Fig_x}, which have the
same scales, for all the $L$ values between $0.16$ m and $0.4$ m and for all
the $R$ values between $\infty $ (perfectly flat sample) and $R=+5$ m
(convex sample) and $R=-5$ m (concave sample). The lens-less case of Fig. \ref{Fig_A1}a
corresponds to the right border of these figures, while the two cases of
Figs. \ref{Fig_A1}b and c are indicated by dashed lines. Fig. \ref{Fig_A2}a maps the distance
among the spots on the light sensor. It shows that that the lens at $L=0.34$%
m shrinks the image by a factor of more than $2$. A band exists around $L=0.3
$ m ($K\simeq F$) in which the measurement is not possible because the spots
on screen cannot be resolved: although the beams are partially focalized,
they still have a finite size. In Fig. \ref{Fig_A2}a, this band is delimited by the
yellow lines, drawn for d $=1$ mm, and the same band is indicated in Figs.
\ref{Fig_A2}b, and \ref{Fig_x}. The limit at 1 mm is not conservative, but the maps are drawn for
d down to 1 mm to appreciate the trends when approaching the measurability
limit. In Fig. \ref{Fig_A2}a the right part of the map presents positive values,
meaning that for $K<F$ the beams impinge on the sensor in the same order they have in
the lens-less case (see Fig. \ref{Fig_A1}b), while the negative values in the left part
indicate that that for $K>F$ they impinge in the reversed order  (see
Fig. \ref{Fig_A1}c). Fig. \ref{Fig_A2}b maps the distance variation d $-$ d$_{\infty }$ (the spot
displacements); positive and negative values indicate the expansion and the
shrinkage of the image. 
From these two maps the map of the relative sensitivity, Fig. \ref{Fig_x}, is derived.
It shows that, perhaps unexpectedly, the behaviours for $K<F$ and for $K>F$
are very different. For $K<F$, the relative sensitivity is always larger than its value in the lens-less case. Starting with the lens very close to the
sensor, and shifting it towards larger distances, the sensitivy starts from
the lens-less values and increases, at first very slowly, then more and more steeply approaching the
region at $K\simeq F$, in which the measurement becomes impossible because
the image is too shrinked. Conversely, when $K$ exceeds $F$, a small
interval of $L$ exists, in which the image is very shrinked and the
sensitivity is high, but then the sensitivity rapidly drops to values much
smaller that those of the lens-less case. Figs. \ref{Fig_A1}b and c allow to
appreciate these different behaviours. The optimal lens position is thus
found for $K<F$. In particular, the distance $L=0.34$ m has been selected,
in which the image is shrinked by a factor of more than 2 (see Fig. \ref{Fig_A2}a),
but remains well readable up to strong curvatures ($R$ down to 5 m or even
less), either convex or concave. For this value of $L$ the relative
sensitivity is larger, by over 20\%, than that of the lens-less case with
the same $A$ (as shown by Fig. \ref{Fig_x}).

\section*{Appendix B: centroid determination algorithm}

\begin{figure}[!t]
\centering
\includegraphics[width = 0.55\columnwidth]{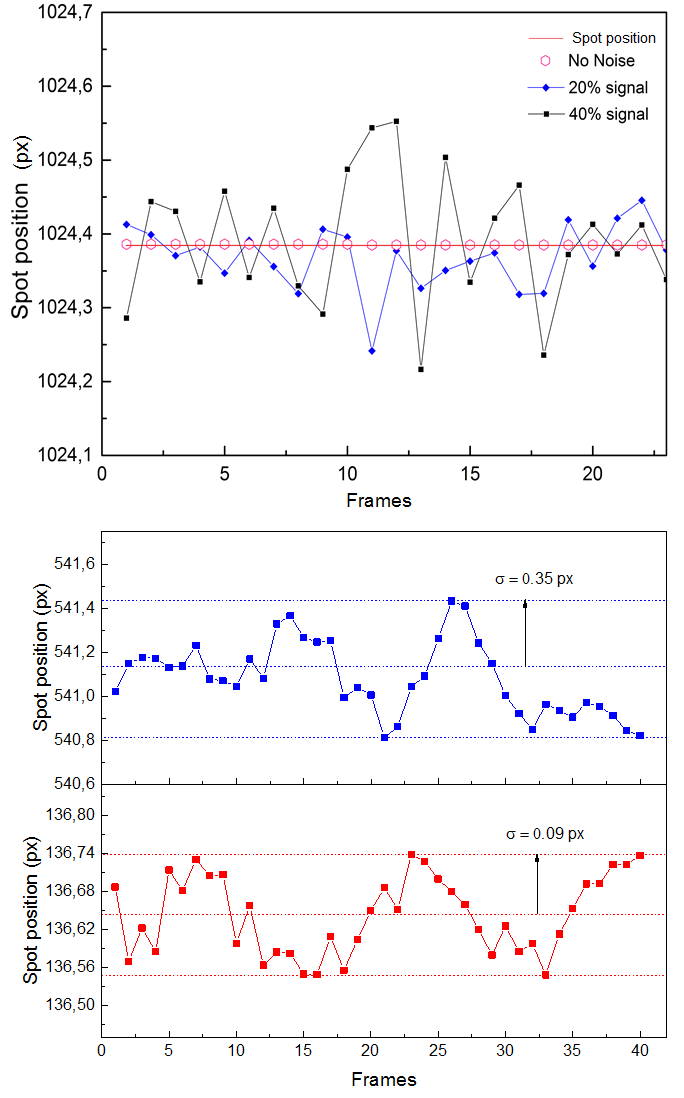}
\caption{a) Single spot centroid determination testing by synthetic code with increasing background noise. b) the blue line refers to a single spot centroid determination in the real apparatus subjected to vacuum system vibration. The red line is the differential spacing between two adjacent spots under the same operating conditions.}
\label{Fig2}
\end{figure}

The algorithm has been synthetically tested to determine the accuracy of the method. The tests are shown in Fig. \ref{Fig2}a in the case of (a) a fixed laser spot, (b) a fixed laser spot with an artificial random background noise of 20$\%$ signal intensity and (c) a laser spot with artificial random background noise of 40$\%$ signal intensity. As it can be expected, the accuracy of the centroid determination is strictly related to the quantity of background noise contained in the image, going from 0.04 pixels to 0.12 pixels in the case of 20$\%$ and 40$\%$ noise respectively. Fig. \ref{Fig2}b shows the repeatability of the determination of the centroids of a fixed laser beam in the real experimental apparatus when the vacuum system is operating (blue line). Here, the accuracy is about 0.3 pixel. If the relative beam spacing (red line) is evaluated instead of the absolute spot position, the accuracy increases up to 0.09 px. With our sensor of 1280 x 1024 pixels, in which the spots are usually at a distance of around 800 pixels, the relative uncertainty $\delta$d$/$d$_\infty$ is of the order of 10$^{-4}$. With a relative sensitivity ($d\left( \text{d}/\text{d}_{\infty
}\right) /d\left( 1/R\right) $) of 1.6 m in the lens-less case and around 2 m in our setup, this corresponds to an uncertainty $\delta (1/R)$ around 5 $\times$ 10$^{-5}$ m$^{-1}$. With a typical curvature radius of 100 m, this means a relative uncertainty $\delta(1/R) / (1/R)$  of 5 $\times$ 10$^{-3}$, down to 5 $\times$ 10$^{-4}$ when the curvature radius decreases to 10 m.  
\end{document}